\documentclass[preprint,aps,prc,showpacs]{revtex4}

\usepackage{graphicx}
\usepackage{bm}

\usepackage{amsmath}
\usepackage{amssymb}
\usepackage[utf8x]{inputenc}
\usepackage{units}
\def \b{\begin{equation}}
\def \e{\end{equation}}
\def \ba{\begin{array}}
\def \ea{\end{array}}
\def \ds{\displaystyle}
\def \be{\begin{eqnarray}}
\def \ee{\end{eqnarray}}

\begin{document}
\title{Strong electric fields induced on a sharp stellar boundary}
\author{Igor N. Mishustin$^{1,2}$, Claudio Ebel$^1$, and Walter Greiner$^1$}
\affiliation{$^1$Frankfurt Institute for Advanced Studies, J.W. Goethe
University, D--60438 Frankfurt am Main, Germany\\ 
$^2$ The Kurchatov Institute, Russian Research Center, 123182 Moscow, Russia}

\date{\today}

\begin{abstract}
Due to a first order phase transition, a compact star may have a discontinuous distribution of baryon as well as electric charge densities, as e.g. at the surface of a strange quark star. The induced separation of positive and negative charges  may lead to generation of supercritical electric fields in the vicinity of such a discontinuity. We study this effect within a relativistic  Thomas-Fermi approximation and demonstrate that the strength of the electric field depends strongly on the degree of sharpness of the surface. The influence of strong electric fields on the stability of compact stars is discussed. It is demonstrated that stable configurations appear only when the counter-pressure of degenerate fermions is taken into consideration.
\end{abstract}


\maketitle

Physics of compact stars is under intensive investigation for many decades. Modern studies are focused on the equation of state (EOS) of dense baryonic matter including possible phase transitions to the quark matter. One interesting feature of a strong first order phase transition is that it can generate a sharp discontinuity in the baryon density as a function of radial coordinate, see e.g. ref. \cite{Bat09}. In the extreme situation when the EOS has a zero pressure point at a finite baryon density $\rho_c$, as is the case e.g. for the MIT bag and NJL models \cite{Dey,NJL}, the discontinuity in the baryon density occurs at the surface of the star, i.e. $\rho(R)=\rho_c$. Such a situation is expected in compact stars made of Strange Quark Matter (SQM), as first predicted in refs. \cite{Witt,Farh}. If the matter is composed of several species with opposite electric charges and different masses, the presence of a sharp discontinuity should lead to a charge separation and generation of an electric field. This effect was first discussed in ref. \cite{Alco} in context of the necked SQM stars, where the electrons from the bulk SQM matter can penetrate in vacuum through the sharp star's surface. As demonstrated in ref. \cite{Pop09}, supercritical electric fields can be generated at the boundary of a nuclear core surrounded by an electronic cloud. In ref. \cite{Neg09} the structure of compact stars with a net charge at the surface was calculated within the General Relativity.   
The boundary effect may be important also for understanding the structure of a mixed phase, where domains of two phases have opposite electric charges \cite{Vosk,Maru}. 
The electrostatic interactions are important for the description of neutron-star crusts where atomic nuclei are embedded in a dense electron gas \cite{Bay71,Bue07}. 

Our goal in the present paper is to study the conditions leading to the generation of supercritical fields at a star boundary. We solve a simplified problem replacing the spherical star boundary by the planar one. We expect this approximation to work very well for a compact star with radius of about 10 km. We assume further that the net positive charge, associated with protons or quarks, has a smooth-step distribution of the Woods-Saxon type
\b \label{step}
\rho_p(z)=\rho_{p0}\left[1+\exp{\left(\frac{z-z_0}{a}\right)}\right]^{-1}~,
\e
where $\rho_{p0}$ is the positive charge density in the bulk ($z\rightarrow-\infty$) and $a$ is a diffuseness parameter. The star boundary is located at $z=z_0$ and $\rho_p(z_0)=\rho_{p0}/2$. At $a\rightarrow 0$ this distribution approaches a rectangular step considered in ref. \cite{Pop09}.

For a macroscopic object like a star the condition of global charge neutrality must hold to a very high precision, see discussion in ref. \cite{Glen}. Therefore, the positive charge of Eq. (\ref{step}) must be fully neutralized by the negative charge of electrons. However, since electrons are light and interact only via the electromagnetic force, they will penetrate through the boundary and generate a local charge disbalance around the star surface. The induced electrostatic potential $\phi(z)$ is determined from the Poisson equation (see e.g. \cite{L&L})
\b \label{Poisson} 
\frac{d^2\phi}{dz^2}=-e\left[\rho_p(z)-\rho_e(z)\right]\equiv -e\rho_{\rm ch}(z)~,
\e  
where $e=\sqrt{4\pi\alpha}$=0.3028 is the proton charge and $\alpha$ is the fine-structure constant\footnote{Here and below we use units with $\hbar=c=1$.}. For a given proton distribution, Eq. (\ref{step}), the electron charge distribution $\rho_e(z)$ should be determined self-consistently. For this purpose we use the Thomas-Fermi approximation \cite{Thomas,Fermi}, which should work well for an extended object like a heavy nucleus or star. The relativistic version of this method was considered e.g. in refs. \cite{Mig76,Fer80}. In a semi-classical approximation the electron energy at point $z$ can be written as \b \label{elen}
\epsilon({\bf k},z)=\sqrt{{\bf k}^2+m_e^2}-V(z)~,
\e   
where ${\bf k}$ is its 3-momentum and $-V(z)=-e\phi(z)$ is the potential energy. At zero temperature all electronic states with $k\leq k_F(z)$ are occupied, where $k_F(z)$ is the local Fermi momentum. It is determined from the condition 
\b \label{mu}
\epsilon(k_F(z),z)=\sqrt{k_F^2(z)+m_e^2}-V(z)=\mu={\rm const}~,
\e 
that gives
\b \label{kF}
k_F(z)=\sqrt{\left[\mu+V(z)\right]^2-m_e^2}~.
\e
The local electron density is found by integration over ${\bf k}$:
\b \label{dens}
\rho_e(z)=2\int\limits_0^{k_F(z)}\frac{d^3k}{(2\pi)^3}=
\frac{\left[\left(\mu+V(z)\right)^2-m_e^2\right]^{3/2}}{3\pi^2}~.
\e
By inserting this expression into Eq. (\ref{Poisson}) we obtain a non-linear differential equation for $\phi(z)$. To find its solutions we need to specify boundary conditions.

According to our assumption, the boundary is in direct contact with the vacuum and therefore the  electron density as well as the electric potential must vanish at $z\rightarrow\infty$.  This condition can be fulfilled if $\mu\leq m_e$. For $\mu<m_e$ the electrons are bound to the surface, i. e. a finite energy is needed to extract an electron from the star. This energy is well known in  ordinary metals as the exit work. In this case the electron density should vanish at a certain distance from the surface, $z_*$, which is found from the condition $m_e-V(z_*)=\mu$, see Eq. (\ref{mu}). The value of $\mu$ is determined from the condition of global charge neutrality: 
\b \label{neutrality}
e\int\limits_{-\infty}^{\infty}\left[\rho_p(z)-\rho_e(z)\right]dz\equiv \sigma_{+}+\sigma_{-}=0~,
\e
where $\sigma_{+}$ and $\sigma_{-}$ are the areal densities of positive and negative charges accumulated at $z<z_0$ and $z>z_0$ respectively. Since calculated $\mu$ values are very close to $m_e$, we use $\mu=m_e$ for our estimates.


Another boundary condition follows from the requirement that far inside the star the electron charge density must completely compensate the positive charge density, i.e. $\rho_{\rm ch}(z\rightarrow -\infty)=0$, that means that $k_{Fe}(z\rightarrow -\infty)\equiv k_{F0}=\left(3\pi^2 \rho_{p0}\right)^{1/3}$ and
\b
V(z\rightarrow-\infty)=\sqrt{k_{F0}^2+m_e^2}-\mu=
\sqrt{\left(3\pi^2\rho_{p0}\right)^{2/3}+m_e^2}-\mu~.
\e  
These conditions are sufficient to determine the electron density distribution $\rho_e(z)$. Unfortunately, no analytical solution can be found without further approximations, and therefore numerical methods must be applied. 

\begin{figure}
\includegraphics[width=1\textwidth]{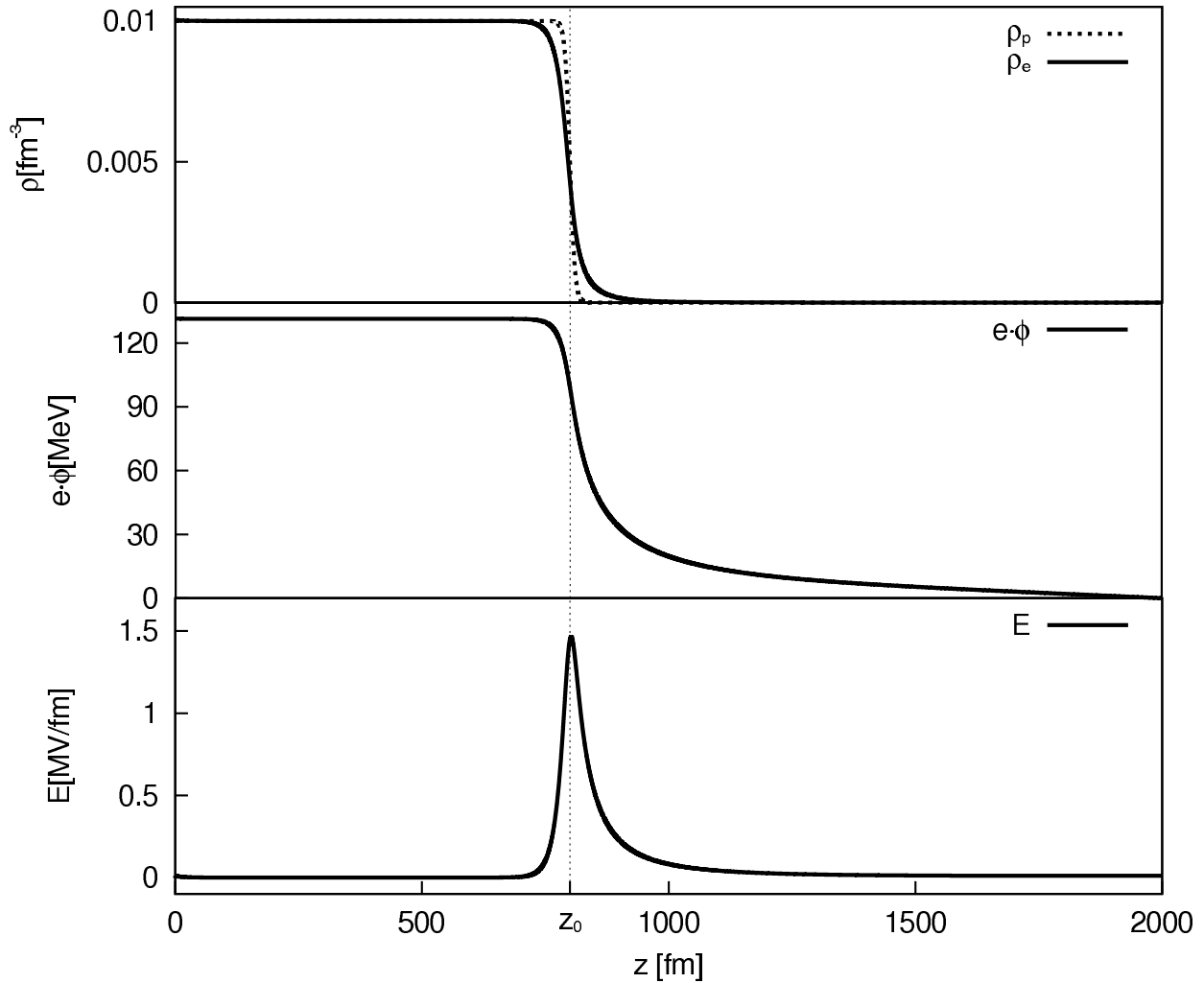}
\caption{The upper, middle and lower panels show, respectively, the electron density (full line), electric potential and electric field, calculated for the given proton density profile (dashed line in the upper panel) as given by Eq.(\ref{step}) with $\rho_{p0}=0.01$ fm$^{-3}$ and $a$=5 fm.}
\end{figure}

We have performed numerical calculations for different values of the diffuseness parameter $a$ ranging from 1 fm, typical for nuclear surfaces, up to 1 nm, typical for solid-state surfaces. As an example, in Fig.~1 we present results for $a$=5 fm which is rather close to a ``sharp surface" limit. The upper, middle and lower panels show, respectively, the electron density, electric potential and electric field calculated for the given proton density (dashed line in the upper panel). One can clearly see the deviation of the electron density from the proton one in the layer of $\pm 50$ fm around the surface. This gives rise to a sharp fall of the potential, and as a consequence, to a very strong electric field generated in this layer. The maximum of the field, about 1.5 MV/fm, is reached at $z=z_0$. This value is almost 600 times higher than the critical electric field for spontaneous electron-positron pair production \cite{Mul}, $E_c=2m^2c^3/e\hbar=0.0026$ MV/fm=$2.6\cdot 10^{16}$ V/cm. However, in the considered system the pair production does not happen because all electron states with $\epsilon_e(k,z)<m_e$ are occupied and therefore Pauli-blocked.

This situation is clarified in Fig. 2, where the electron levels in a strong  electric potential, $-V(z)$, are shown schematically. As explained above, the negative potential inside the medium ($z<z_0$) is generated by the charge separation at the surface. In the vacuum such a potential would certainly lead to a spontaneous electron-positron pair production. In this process the negative-energy electrons from the Dirac sea (below the mass gap) would penetrate through the barrier and occupy a state in the upper continuum (above the mass gap) leaving a hole (positron) behind. But all electron states up to $\epsilon=\mu$ are occupied (Fermi sea) to balance the bulk positive charge of protons. The situation is changed at finite temperatures when the particle-hole excitations are produced in the Fermi sea. Then, the electrons from the Dirac sea can occupy the vacancies in the Fermi sea. The process of thermal emission of electron-positron pairs from the sharp surface was considered earlier in ref. \cite{Usov}. Here we want to point out other interesting processes, namely, the spontaneous production of negative muons and pions, which may replace the electrons. They become possible in a strong enough field when $V_0>m_\mu,~m_\pi$. This opens the possibility of pion condensation induced by strong electric fields \cite{Mig76}.

\begin{figure}
\includegraphics[width=1\textwidth]{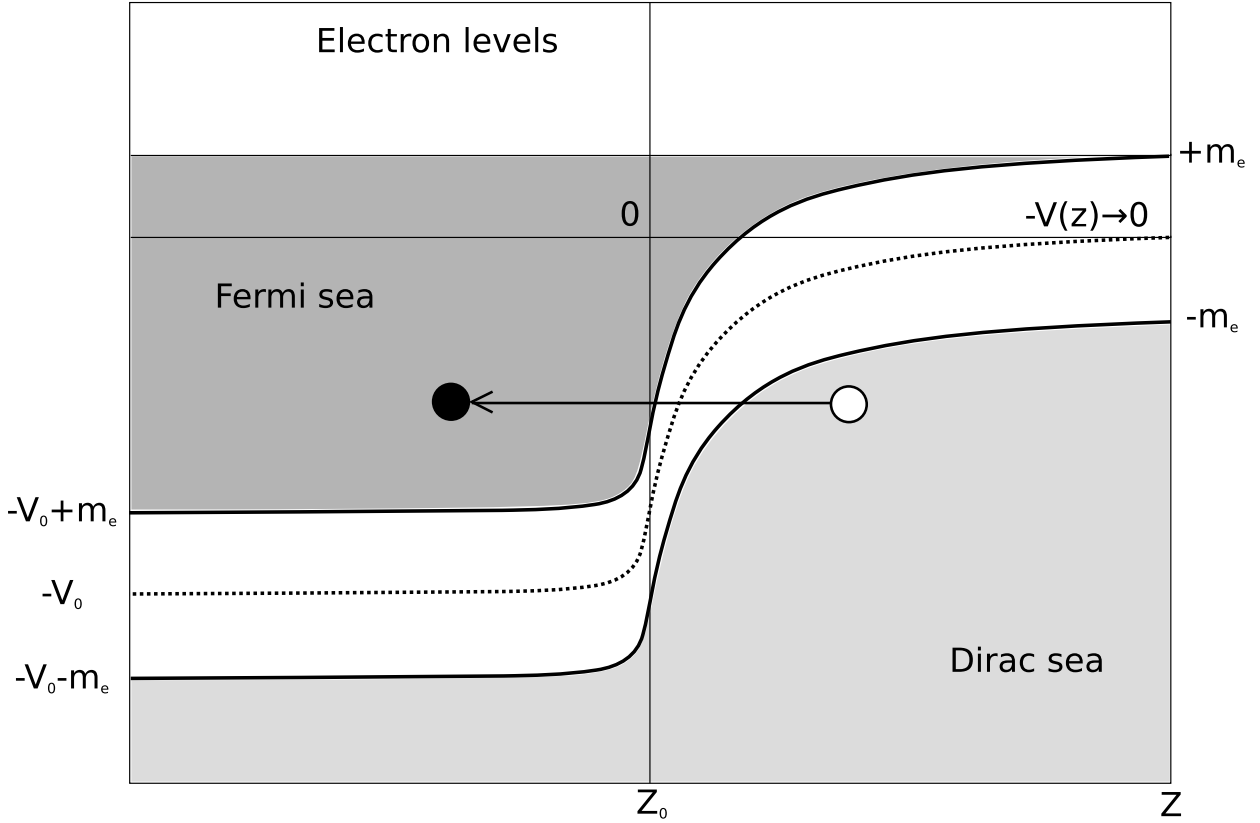}
\caption{Schematic view of the electron energy levels in a strong electrostatic potential $-V(z)$. The electron states above (Fermi sea) and below (Dirac sea) the mass gap (white area) are shown by dark and grey shadowing. Transitions from the Dirac to Fermi sea (long arrow) are not allowed because all states in the Fermi sea are occupied. 
}
\end{figure}

With increasing $a$ the electron density comes closer and closer to the proton density and the charge separation at the surface diminishes. The electric potential difference at   $z\rightarrow\pm\infty$ is fixed by the boundary conditions, but its gradient, i. e. the electric field, across the transition region becomes smaller and smaller. We have checked that the electric field falls below the critical value at $a\approx 10$ pm. In Table 1 we give the maximum value of the electric field, E$_{\rm max}$, calculated for several values of the diffuseness parameter $a$.
E$_{\rm max}$ can be calculated by using the formula which follows from the Gauss law:
\b \label{Gauss}
{\rm E}_{\rm max}\equiv {\rm E}(z=z_0)=e\int\limits_{-\infty}^{z_0}\rho_{\rm ch}(z)dz=\sigma_{+}~,
\e
where $\sigma_{+}$ is the areal density of the net positive charge at $z<z_0$.

\begin{table}
\caption{Maximum electric field and areal charge density calculated for several values of the diffuseness parameter $a$.}
\begin{center}
\fbox{
\begin{tabular}{r|c|l|l}
\multicolumn{1}{c|}{ \boldmath
$a$
} &

\multicolumn{1}{c|}{ \boldmath
$E_{max}\left[ \unitfrac{MV}{fm} \right]$
} &

\multicolumn{1}{c|}{ \boldmath
$\nicefrac{E_{max}}{E_c}$
} &
\multicolumn{1}{c}{ \boldmath
$\sigma_{+}\left[e/fm^2\right]$
}\\
\hline
\unit[1]{fm} & $1.82 \cdot 10^{-0}$ & $6.89 \cdot 10^{+2}$ & $1.01 \cdot 10^{-1}$ \\
\unit[10]{fm} & $1.16 \cdot 10^{-0}$ & $4.38 \cdot 10^{+2}$ & $6.41 \cdot 10^{-2}$ \\
\unit[100]{fm} & $2.05 \cdot 10^{-1}$ & $7.74 \cdot 10^{+1}$ & $1.13 \cdot 10^{-2}$ \\
\unit[1]{pm} & $2.08 \cdot 10^{-2}$ & $7.86 \cdot 10^{~0}$ & $1.18 \cdot 10^{-3}$ \\
\unit[10]{pm} & $2.08 \cdot 10^{-3}$ & $7.86 \cdot 10^{-1}$ & $1.18 \cdot 10^{-4}$ \\
\unit[100]{pm} & $2.08 \cdot 10^{-4}$ & $7.86 \cdot 10^{-2}$ & $1.18 \cdot 10^{-5}$ \\
\unit[1]{nm} & $2.08 \cdot 10^{-5}$ & $7.87 \cdot 10^{-3}$ & $1.25 \cdot 10^{-6}$ \\


\end{tabular}
}	
\end{center}
\end{table}

For our discussion below we need to estimate the electric field and other related quantities on a qualitative level. This can be done by using the ultrarelativistic limit, $V\gg m_e$, which is justified everywhere except of the asymptotic region at $z\rightarrow\infty$. Taking for simplicity the limit $a\rightarrow 0$, we get the exact solution of the Poisson equation (\ref{Poisson})\footnote{Some analytical solutions of the relativistic Thomas-Fermi equation for the case of spherical symmetry were found earlier in refs. \cite{Pop09,Rot09}.}
\be \label{potential}
V(z)=\left\{\begin{array}{ll}
V_0\left[1-c_1 \exp{\left(\frac{\ds z-z_0}{\ds\lambda}\right)}\right], &z<z_0\\
V_0\left[c_2+\frac{\ds (z-z_0)}{\ds\sqrt{6}\lambda}\right]^{-1}, &z>z_0 \end{array}\right.
\ee
where 
\b
\lambda=\frac{\pi}{eV_0}\approx \frac{\pi}{ek_{F0}}~.
\e 
The constants $c_1=0.2374$ and $c_2=1/(1-c_1)=1.3113$ are obtained from the continuity conditions for $V(z)$ and $dV/dz$ at $z=z_0$. It is interesting that the decay length $\lambda$ is determined by the electron Fermi momentum in the bulk of the medium $k_{F0}$. One can see that the potential changes much faster at $z<z_0$ than at $z>z_0$. 
 The electric field can be easily obtained now by differentiating Eq. (\ref{potential}),
\be
{\rm E}(z)=\left\{\begin{array}{ll}
c_1\frac{\ds V_0}{\ds e\lambda}\exp{\left(\frac{\ds z-z_0}{\ds\lambda}\right)}, &z<z_0\\
\frac{\ds V_0}{\ds \sqrt{6}e\lambda}\left[c_2+\frac{\ds (z-z_0)}{\ds\sqrt{6}\lambda}\right]^{-2} &z>z_0 \end{array}\right.
\ee
It is obvious that the maximum of the electric field is reached at $z=z_0$: 
\b
{\rm E}_{\rm max}=c_1\frac{V_0}{e\lambda}=c_1\frac{V_0^2}{\pi}=c_1\frac{k^2_{F0}}{\pi}~.
\e
The areal charge density at $z=z_0$ is given by eq. (\ref{Gauss}). Assuming that this charge is distributed over the layer of width $\lambda$ we can estimate the average charge density in this layer  
\b
e\tilde{\rho}_{\rm ch}\approx \frac{\sigma_{+}}{\lambda}=c_1\frac{ek^3_{F0}}{\pi^2}~,
\e
i. e. about 70\% of the proton density at $z\rightarrow -\infty$.
The energy density of the electric field is ${\cal E}_{\rm em}={\rm E}^2/2$, and therefore the total electrostatic energy is easily calculated as
\b \label{em}
E_{\rm em}=\int\limits_{-\infty}^{\infty}{\cal E}_{\rm em}dV=\frac{c_1(2-c_1)}{12}\frac{V_0^2}{e^2\lambda}S\approx 
c_1(2-c_1)\frac{\pi S}{4e}\rho_p~,
\e
where $S$ is the surface area and $\rho_p$ is the bulk proton density. Note that this expression is approximately equal to the energy stored in a planar capacitor with areal charge density $\sigma_{+}$ and gap $\lambda$.

The authors of ref. \cite{Pop09} discuss the possibility of a new family of stable star-like objects (nuclear cores) where the repulsive electromagnetic force, associated with the charged layer at a sharp boundary, is balanced by the gravitational force. Here we present our counter-arguments concerning such possibility. First of all we point out that in a macroscopic nuclear matter, besides neutrons and protons, the electrons must be present to neutralize the bulk proton charge. By this reason the $\beta$-equilibrium will be shifted to the neutron-rich side, so that the matter will contain mainly neutrons with a small admixture of protons and electrons. According to standard calculations of $\beta$-equilibrated npe-matter in neutron stars (see e.g. refs. \cite{Bay71,Glen}), the fraction of protons, $\xi=\rho_p/\rho_B$ should be on the level of a few percent. As well known, such matter is unbound, in contrast to the symmetric nuclear matter with $N_n\approx N_p$ without electrons.

For simplicity we replace the baryon, proton and neutron densities by their mean values:
\b \label{rho}
\rho_B=\frac{3N_B}{4\pi R^3}~,~~~\rho_p=\frac{k_{Fp}^3}{3\pi^2}=
\xi\rho_B~,~~~\rho_n=\frac{k_{Fn}^3}{3\pi^2}=
(1-\xi)\rho_B~,
\e
where $R$ is the star radius. 


Now let us consider a star-like configuration with sharp boundary at $r=R$, where we expect formation of a strong electric field. In our qualitative discussion we represent the star energy per baryon as the sum of three contributions:
\b \label{k-e-g}
W(R)=W_{\rm kin}(R)+W_{\rm em}(R)+W_{\rm grav}(R)~,
\e
where terms in the r.h.s. represent the kinetic energy of degenerate fermions (npe), the electrostatic energy and the gravitational energy. The kinetic energy term is calculated easily for any fermionic species of mass $m_i$ and Fermi momentum $k_{Fi}$:
\be 
\label{kin}
W^{(i)}_{\rm kin}(x_i)=\frac{3m_i}{8x_i^3}\left[\left(2x_i^3+x_i\right)\sqrt{1+x_i^2}-{\rm arcsinh}\left(x_i\right)\right]
=\left\{\begin{array}{ll}
m_i\left(1+\frac{3}{10}x_i^2\right),~~x_i\ll 1 \label{kin-a} \\ 
\frac{3}{4}x_i,~~x_i\gg 1 \label{kin-b}
\end{array}\right.~
\ee
where i=n,p,e and $x_i=k_{Fi}/m_i$. Below we include in $W_{\rm kin}$ only neutron contribution and express $k_{Fn}$ as a function of $R$ using Eq. (\ref{rho}),  
\b
k_{Fn}=\left(\frac{9\pi}{4}\right)^{1/3}\frac{N_n^{1/3}}{R}~.
\e
 
The electrostatic energy of a spherical dipole layer can be obtained from Eq. (\ref{em}) by replacing $\rho_p=3\xi N_B/4\pi R^3$ and $S=4\pi R^2$. Then the energy per baryon is 
\b
W_{\rm em}(R)=\frac{3\pi}{4}c_1(2-c_1)\frac{\xi}{eR}\equiv c_3\frac{\xi}{eR}~,
\e
where $c_3\approx 0.426$. It should be emphasized that this energy is by factor $\frac{\lambda}{R}\sim 1/N_p^{1/3}$ smaller than the energy of net charge $Q=4\pi R^2\sigma_{+}$ distributed in a spherical layer of wdth $\lambda$ and radius $R$ (see ref. \cite{Neg09}).  
Finally, the gravitational energy of a baryon at the surface of  a star of mass $M=N_Bm_B$ and radius $R$ is easily calculated in Newtonian gravity:
\b
W_{\rm grav}(R)=-\frac{GMm_B}{R}=-\left(\frac{m_B}{m_{\rm Pl}}\right)^2\frac{N_B}{R}~,
\e
where $m_{\rm Pl}=1/\sqrt{G}\approx 1.22\cdot 10^{19}$ GeV is the Planck mass and $m_B=939$ MeV is the nucleon mass.

Putting together all the contributions we can now rewrite Eq. (\ref{k-e-g}) as
\b \label{energy}
W(R)=\frac{3}{4}\left(\frac{9\pi}{4}\right)^{1/3}\frac{N_n^{1/3}}{R}+
\left[c_3\frac{\xi}{e}-\left(\frac{m_B}{m_{\rm Pl}}\right)^2N_B\right]\frac{1}{R}~,
\e
where the ultrarelativistic limit of Eq. (\ref{kin}) has been used in the kinetic term. One can immediately see that the electrostatic and gravitational contributions become equal at 
\b
N_{\rm min}=c_3\frac{\xi}{e}\left(\frac{m_{\rm Pl}}{m_{B}}\right)^2\approx 6.26\cdot 10^{36}~,
\e 
where the value $\xi=1/38$ was used for the estimation. The corresponding mass $M_{\rm min}=N_{\rm min}m_B$ is about $10^{13}$ g, i. e. almost 20 orders of magnitude smaller than the maximum mass of a neutron star. At $N_B<N_{\rm min}$ there exists no equilibrium configurations because the kinetic pressure will force the system to expand indefinitely. 

At $N_{B}>N_{\rm min}$ the negative gravitational contribution exceeds the electrostatic one. The latter becomes completely negligible and can be neglected at baryon numbers $\sim 10^{57}$ characteristic for compact stars. As was first demonstrated by Chandraseckhar \cite{Chan} and Landau \cite{Land}, a compact star can be stabilized by the pressure of a degenerate fermion gas alone. 
From (\ref{energy}) one can see that the balance between kinetic and gravitational energy can be achieved if the total number of baryons is smaller than  
\b
N_{\rm max}=\left(\frac{3}{4}\right)^{3/2}\left(\frac{9\pi}{4}\right)^{1/2}\left(\frac{m_{\rm Pl}}{m_n}\right)^3\approx 3.78\cdot 10^{57}~.
\e 
As follows from this simple calculations, the corresponding maximum mass of a neutron star, $M_{\rm max}=N_{\rm max}m_n$ is about 3.2 $M_{\odot}$. Numerical calculations within General Relativity \cite{TOV} for the degenerate noninteracting neutron gas predict somewhat lower maximum mass, $M_{\rm max}=0.7M_{\odot}$. The repulsive neutron-neutron interaction may  
increase this value up to $4M_{\odot}$ (see e.g. ref.\cite{Nara}). It is easy to see that stars with $N_{\rm min}<N_B<N_{\rm max}$ have a stable equilibrium state. Indeed,  Eq. (\ref{energy}) shows that in this case $W(R)>0$ and changes as $1/R$, so that the star should expand to lower densities. But this expansion will terminate at a large enough $R$, when the neutron gas becomes non-relativistic. As follows from Eq. (\ref{kin}), in this limit the energy per particle is $m_n+3k^2_{Fn}/10m_n$, i. e. changes as $1/R^2$ at large $R$. Since the negative gravitational contribution still changes as $1/R$, it wins so that the total energy per baryon approaches $m_n$ from below. Therefore a minimum develops in $W(R)$. This is famous Landau's argument presented in ref. \cite{Land}. In ref. \cite{Nara} it was reformulated for the case of General Relativity and for an arbitrary fermion mass.

It is clear now that ignoring the kinetic-energy term in Eq. (\ref{energy}), as done in ref. \cite{Pop09}, will lead to the erroneous conclusion regarding the star stability. Although the electrostatic and gravitational forces can balance each other at $N_B=N_{\rm min}$, this state corresponds to a maximum of energy. As one can easily see,    the system will either expand or collapse depending on its actual baryon number compared to $N_{\rm min}$. Moreover, the electrostatic force can support only the surface layer of matter where the electric field is nonzero. The inner layers of the star require the kinetic pressure to resist gravity. Therefore, the mechanism discussed in ref. \cite{Pop09} cannot produce any new stable object. Inclusion of the electrostatic effects will only slightly increase the star mass, see e. g. ref. \cite{Neg09}. 

In conclusion, we have demonstrated that very strong electric fields can be generated on a sharp boundary separating the bulk matter from the vacuum. The strength of the field is determined by the separation of positive and negative charges in the surface layer. For the proton charge distribution with the diffuseness parameter of a nuclear scale the field strength may exceed significantly the critical value for the electron-positron pair production. The field strength diminishes when the charge distribution becomes more and more smooth. We have also considered star-like configurations with a strong electric field on the boundary. Our simple analysis shows that such electric field alone cannot balance the gravitational force to produce stable objects. The stable compact star configurations appear only when the counter-pressure of degenerate fermions is taken into consideration.  For a more accurate description of compact stars with strong electric fields one should perform calculations within General Relativity and include effects of strong interactions in the energy of the nuclear core.

We are grateful to Prof. Remo Ruffini who attracted our attention to the problems discussed in this paper. We acknowledge his kind hospitality during our visit to the International Center for Relativistic Astrophysics (ICRA) at Pescara in March 2009. We also thank Dr. Thomas B\"urvenich and members of the ICRA group for fruitful discussions. This work was supported in part by the DFG Grant 436 RUS 113/957/0-1 (Germany), and by the Grants NS-3004.2008.2 and RFBR 09-02-91331 (Russia).

\end{document}